\documentclass[a4paper,fleqn,usenatbib]{mnras}
\usepackage[T1]{fontenc}
\usepackage{ae,aecompl}
\usepackage{graphicx}	
\usepackage{amsmath}	
\usepackage{amssymb}	
\usepackage{enumerate}
\usepackage{color}

\title[Radius determination of open clusters]{A method for
       determining the radius of an open cluster from stellar
       proper motions}

\author[S\'anchez et al.]{
N\'estor S\'anchez$^{1}$\thanks{E-mail: nestor@um.es (NS)},
Emilio J. Alfaro$^{2}$ and F\'atima L\'opez-Mart\'inez$^{3}$\\
$^{1}$Departamento de F\'{\i}sica, Universidad de Murcia,
E-30100 Murcia, Spain.\\
$^{2}$Instituto de Astrof\'{\i}sica de Andaluc\'{\i}a, CSIC,
Glorieta de la Astronom\'{\i}a s/n, 18008, Granada, Spain.\\
$^{3}$Centro de Astrof\'isica da Universidade do Porto,
Rua das Estrelas, P-4150-762 Porto, Portugal.}

\date{Accepted XXX. Received YYY; in original form ZZZ}
\pubyear{2017}

\begin{document}
\label{firstpage}
\pagerange{\pageref{firstpage}--\pageref{lastpage}}
\maketitle

\begin{abstract}
We propose a method for calculating the radius of an open
cluster in an objective way from an astrometric catalogue
containing, at least, positions and proper motions. It uses
the minimum spanning tree (hereinafter MST) in the proper
motion space to discriminate cluster stars from field stars
and it quantifies the strength of the cluster-field separation
by means of a statistical parameter defined for the first time
in this paper. This is done for a range of different sampling
radii from where the cluster radius is obtained as the size
at which the best cluster-field separation is achieved.
The novelty of this strategy is that the cluster radius
is obtained independently of how its stars are spatially
distributed. We test the reliability and robustness of
the method with both simulated and real data
from a well-studied open cluster (NGC~188),
and apply it to UCAC4 data for five other
open clusters with different catalogued radius values.
NGC~188, NGC~1647, NGC~6603 and
Ruprecht~155 yielded unambiguous radius values
of $15.2 \pm 1.8$, $29.4 \pm 3.4$, $4.2 \pm 1.7$
and $7.0 \pm 0.3$~arcmin, respectively. ASCC~19 and
Collinder~471 showed more than one
possible solution but it is not possible to know whether
this is due to the involved uncertainties or to the presence
of complex patterns in their proper motion distributions,
something that could be inherent to the physical object
or due to the way in which the catalogue was sampled.
\end{abstract}


\section{Introduction}
\label{secintro}

Star clusters have long been recognized as very useful
tools in many areas of astronomy including, among others,
the structure and evolution of the Milky Way
\citep[see for instance][]{Gil12,Ran13,Mor16}.
A precise knowledge of cluster properties such as
distance, age, metallicity or reddening is necessary
in order to be able to draw reliable conclusions.
Large open cluster catalogues, like that published by
\citet{Dia02,Dia14}, compile all the available information
required to make studies on, for instance, the rotation
of the spiral patterns \citep{Dia05} or the Galactic star
formation history \citep{Fue04}. However, this kind of
data collections has the disadvantage of being highly 
heterogeneous. With the increasing number of publicly
available photometric and astrometric databases,
there is a growing interest in the automated and
systematic estimation of homogeneous parameters for
Galactic cluster
\citep[e.g.][]{Kha12,Dia14,Kro14,Sar14,Per15,Sam17}.
The catalogue by \citet{Kha13}, mainly based on the
PPMXL catalogue \citep{Roe10}, provides basic
astrophysical data for a large set of clusters
derived in a uniform and homogeneous way. There
is however a need for some caution in this kind of
massive data processing because slight variations
in the developed strategies can lead to significant
biases in the inferred cluster parameters.
\citet{Net15} compared the parameters given in several
catalogues \citep[including][]{Kha13} and concluded
that there are clear discrepancies and trends in
distances, reddenings and ages.

Cluster radius is a particularly valuable parameter
because it is a common strategy to choose the size of
the field of view surrounding the cluster very close
to cluster size in order to minimize contamination by
field stars. In fact, \citet{San10} have shown that,
when estimating cluster memberships for a
mixture of two Gaussian distributions,
a sampling radius larger than the cluster radius may
produce a severe contamination by field stars in the
identified cluster members (spurious members) that
certainly affects the determination of the remaining
cluster properties. Apart from visual inspection, the 
standard method for directly estimating cluster radii
is based on their projected radial density profiles.
Usually, a King-like function (or any other analytical
function) is fitted to the density profile and the
cluster radius is extracted from this
fit. Systematic determinations of cluster sizes
based on this strategy have been performed by
\citet{Kha05a,Kha12,Kha13} and \citet{Pis07,Pis08}
and compiled in their final catalogue of cluster
parameters \citep{Kha13}. However,
\citet{Kha05a} pointed out that their published
radii are $\sim 1.5-2.5$ times larger than
the corresponding values compiled by \citet{Dia02}.
The last re-calculation of cluster radii for
all the $2167$ clusters listed in \citet{Dia02} was
made by \citet{Sam17} and their results agree 
reasonably well with those by \citet{Dia02}.
The main limitation of the radial density method is
its sensitivity to small variations in the distribution
of stars, especially for poorly populated open clusters.
Moreover, this kind of strategy is not appropriate for
open clusters exhibiting a high degree of substructure
\citep{San09}.

In this work we propose a different approach to the
problem. The idea is based on our previous result
that, for normally distributed proper motions,
the best sampling radius, i.e. the sampling that best
separates cluster and field populations
(based on a Gaussian-mixed model fitting),
very closely
coincides with the ``true" cluster radius \citep{San10}.
We use proper motions to discriminate cluster stars from
field stars and quantify the quality of the cluster-field
separation. We do this for a range of different sampling 
radii from which we obtain the cluster radius.
In order to separate cluster from field stars in the
proper motion space we do not use the standard method
of fitting two Gaussian functions \citep{Vas58,San71,Cab85}
because contamination by field stars at large sampling
radius may yield unrealistic results \citep{San10}.
Instead, we use a methodology based on the minimum
spanning tree (MST) of the stars. The MST
is the set of straight lines connecting the 
points such that the sum of their lengths is minimum.
MST clustering algorithms are known
to be capable of detecting clusters with irregular
boundaries and have been used in astronomy for
searching and characterizing large-scale structures
\citep{Bar85,Wan16}, stellar systems
\citep{Car04,Schm06,Koe08,Sch08,Gut09,San09,Gre15,Alf16,Beu17,Dib17,Jaf17}
and even interstellar clouds \citep{Car06,Lom11}.
It is important to note that here
we are not searching for or characterizing
open clusters. We assume there is actually a
cluster and we use the spanning tree to delimit
the cluster overdensity in the proper motion
space and from there determine what its radius
is. Thus, the procedure does not use positions
but proper motions without any parametric model
assumption, which allow us to calculate the radius
in an objective way independently of how cluster 
stars are spatially distributed.

In Section~\ref{secmetodo} we describe
in detail the proposed method. We first simulate a
well-behaved, homogeneous cluster to explain how the
method works (Section~\ref{idealcase}) and to define
what we call the transition parameter (Section~\ref{seceta}).
Some tests on simulated Gaussian distributions are shown
in Section~\ref{secsimula}. The strategies for estimating
the cluster radii and the uncertainties are described in
Sections~\ref{secradio} and \ref{secerrores}, respectively.
Additionally, we use the well-studied open cluster NGC~188 
as a test case for validating the reliability of the method
(Section~\ref{secngc188}).
Section~\ref{secsample} gives the sample of selected open clusters
whose radii are estimated and discussed in Section~\ref{seccumulos}.
Finally, in Section~\ref{conclusion} we summarize the main findings.

\section{Method}
\label{secmetodo}

\begin{figure*}
\includegraphics[width=0.66\columnwidth]{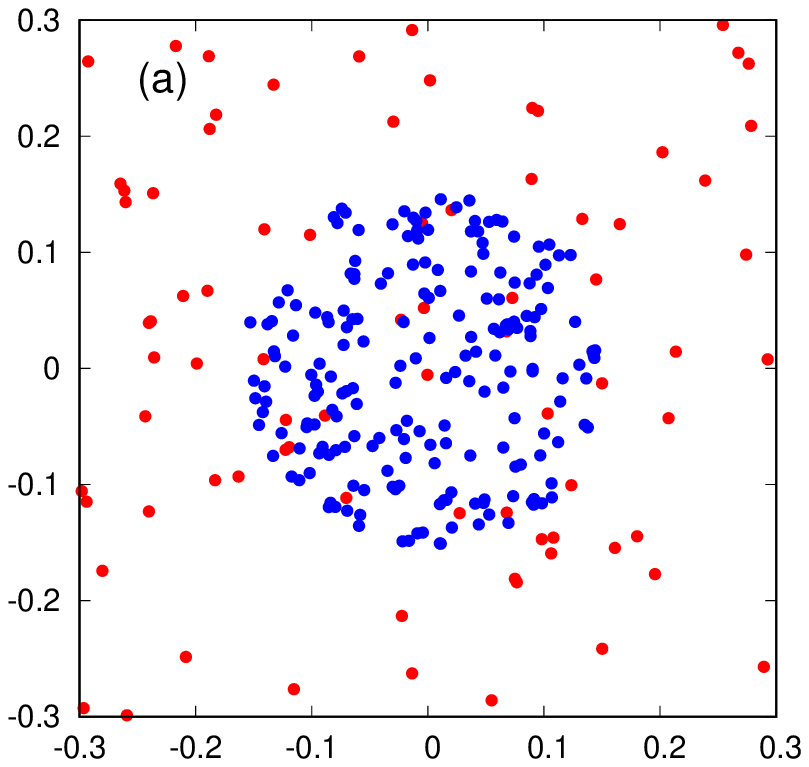}
\includegraphics[width=0.66\columnwidth]{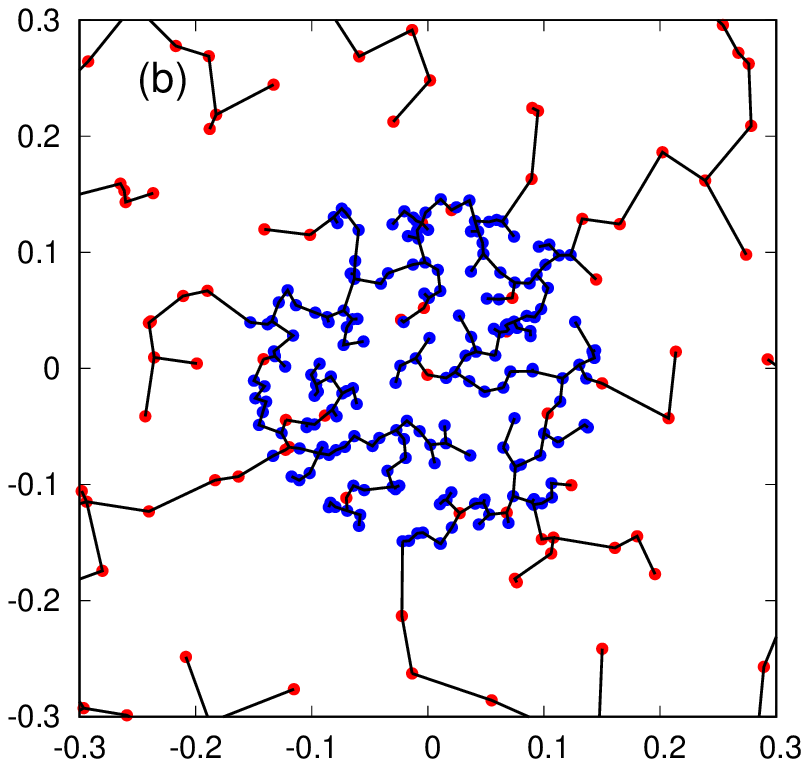}
\includegraphics[width=0.66\columnwidth]{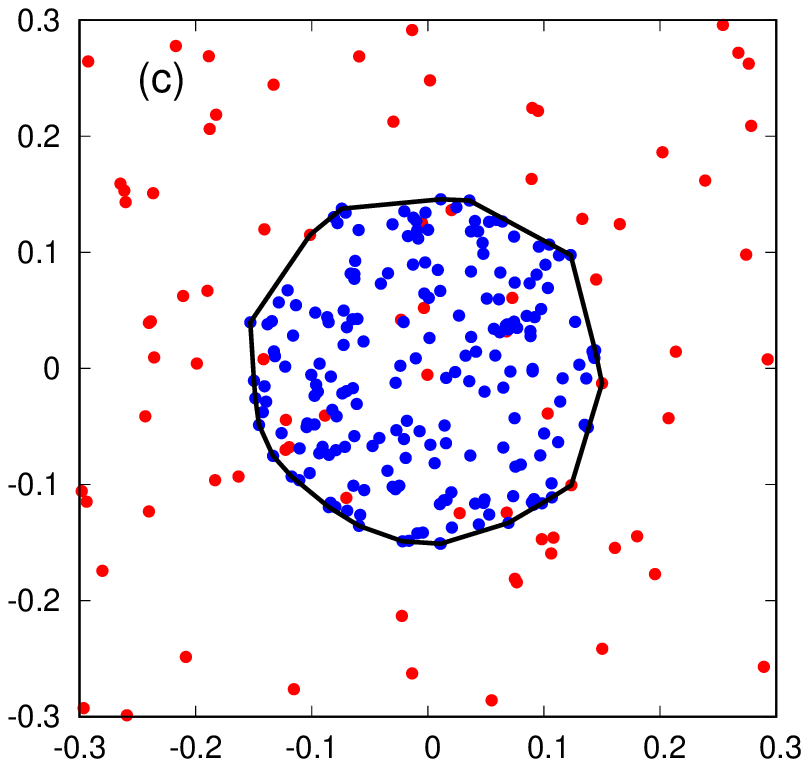}
\caption{Simulation of $200$ cluster stars and $800$
field stars homogeneously distributed in a circular area
of radius $1.0$, such as the cluster-field surface density
ratio is $10$. (a) Distribution of stars in the
central region: blue points are cluster stars and red
points field stars. (b) The corresponding MST of the
central region of the data sample. (c) The plotted convex
hull surrounds the data points that the algorithm classifies
as belonging to the overdensity
(details in the text).}
\label{simulaxy}
\end{figure*}

We adopt the working definition of a star cluster as an overdensity
in a given phase-space diagram. Ideally, some kind of clustered
structure should be seen for the full set of phase-space variables
(positions, parallaxes, proper motions and radial velocities) but
this is not always the case either because some of these variables
are not available or because contamination by field stars hides
the underlying clustered structure in some subspace.
Here we consider only one set of two variables (proper motion).
Let us assume there is a cluster inside a more spread distribution
of field stars. The branch lengths of the MST constructed from the
whole set of points should exhibit some kind of bimodal distribution
with small branches corresponding to connections in the region of
the diagram occupied by the cluster and large branches corresponding
to connections among field stars. That is, the mean of the cluster
branches should be, on average, smaller than the mean of the field
branches.

For constructing the MST we use the Prim's algorithm \citep{Pri57}.
At a given iterative step, we search for and add the smallest branch
connecting points that are not part of the MST with points that are
already part of the tree. If the starting point is a star in the
cluster, the algorithm first adds to the tree stars belonging to
the cluster (i.e. that are located in the high density region)
because those stars are separated by the smallest distances.
In an ideally-behaved case, field stars will be added to the
MST only when all the cluster stars have been already included.
At each step of the Prim's algorithm, for the $N_T$ points that 
are part of the tree we calculate the mean length of the branches
($L_T$). If, for instance, we start from a star in the region
covered by a nearly homogeneous cluster, we expect that $L_T$
remains approximately constant when we add cluster stars to
the tree and it starts to increase when field stars with
larger separations are included. This is the property we
take advantage of to separate cluster from field.

A key point is the starting point. We are proceeding
under the assumption that there is actually a cluster
in the data sample.
Our goal in this work is to derive the cluster radius
and not to decide whether there is or not a cluster.
The starting point can be set up by hand if the
cluster position in the phase-space diagram is
known. However, as we plan to apply this method
massively and systematically to data from Gaia
mission, we included a function in our code
to automatically select as starting point the
densest part of the tree, i.e. that with the
maximum number of stars per unit length. To
calculate the {\it local} density we consider a
subsample of $N_{min}$ data points (see below
the assumed value of this parameter). In any
case, our tests showed that the method works
well for any starting point as long as it
is inside or close to the region occupied by the
cluster.

\subsection{Homogeneous cluster case}
\label{idealcase}

\begin{figure}
\includegraphics[width=\columnwidth]{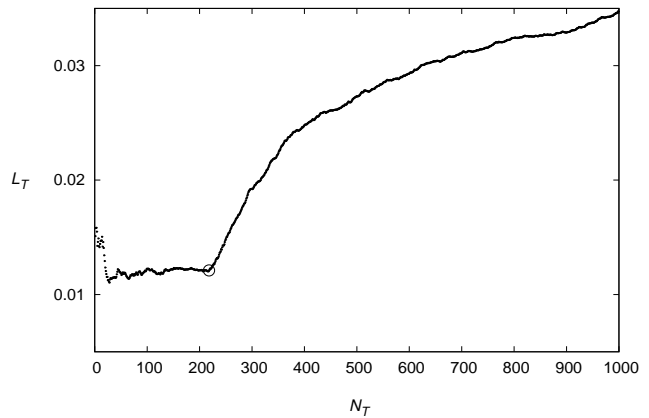}
\caption{Mean length of the branches ($L_T$) as a function
of the number of points that are part of the MST ($N_T$)
for the simulated data shown in Fig.~\ref{simulaxy}.
Open circle indicates the point where the transition
cluster-field occurs (see text).}
\label{simulamedia}
\end{figure}

In order to understand how the method works, it is
useful to see the results for a well-behaved cluster.
For this we simulate two homogeneous distributions
one of which is denser than the other (overdensity).
Obviously this does not correspond to the case of
two nearly gaussian distributions that we would
expect in the proper motions space (we will show
these simulations in Section~\ref{secsimula}),
but this simple
example case will serve to illustrate the main
features and performance of the proposed method.
The simulation consists of $1000$ stars randomly
distributed in a circle
of radius $1.0$ (arbitrary units), from which
$200$ are cluster stars that are distributed
in a denser region. The cluster-field surface
density ratio is $10$.
It is important to mention that for this simple
ideal case the area covered by the overdensity 
in this phase-space overlaps with the cluster
itself, but this will not be the case for more
realistic clusters having radial density
distributions
(Section~\ref{secsimula}). In any case, some
field stars are
located by chance below the area occupied by
cluster stars (Fig.~\ref{simulaxy}a).
Unless we use additional information from other physical
variables, these stars will be incorrectly classified
as cluster stars by this and any other method. If we
construct the MST (Fig.~\ref{simulaxy}b) starting from
a cluster star and we plot mean length of the branches at
each step we get what is shown in Fig.~\ref{simulamedia}.
At the beginning of constructing the MST we see some
statistical fluctuations for low values of $N_T$, but
after that $L_T$ remains fairly constant around the
average separation of cluster stars in the phase-space
diagram. After including all the $200$ cluster stars
(plus some additional field stars below the cluster),
$L_T$ begins to increase as new longer branches
corresponding to the field are added to the MST.
The transition from cluster to field is easily
visible in Fig.~\ref{simulamedia} and it is the
key property we use to separate cluster from field.
We consider as cluster all the stars that are part
of the MST at the transition point in the $L_T-N_T$
plot (open circle in Fig.~\ref{simulamedia}).
The convex hull\footnote{The convex hull is the
minimum-area convex polygon containing the set
of data points.} containing these points 
(Fig.~\ref{simulaxy}c) shows that the selection
is done properly.
\begin{figure*}
\includegraphics[width=\columnwidth]{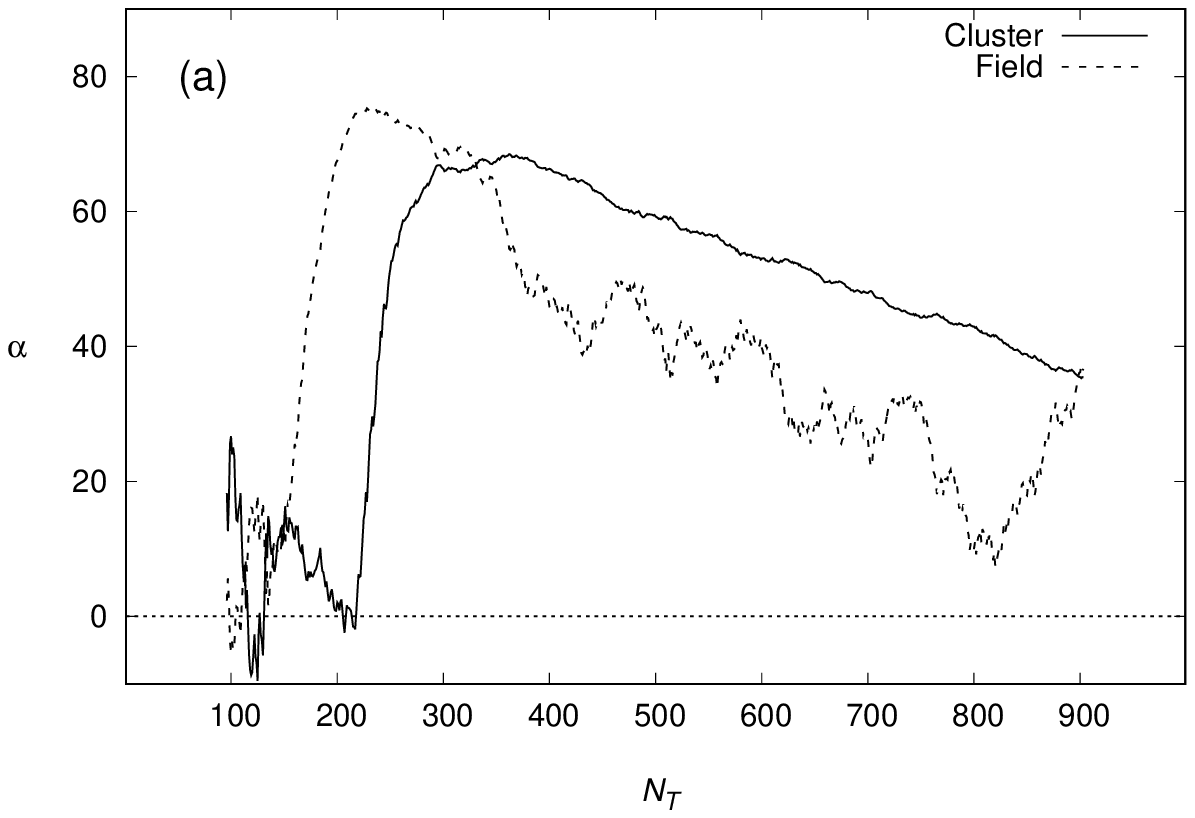}
\includegraphics[width=\columnwidth]{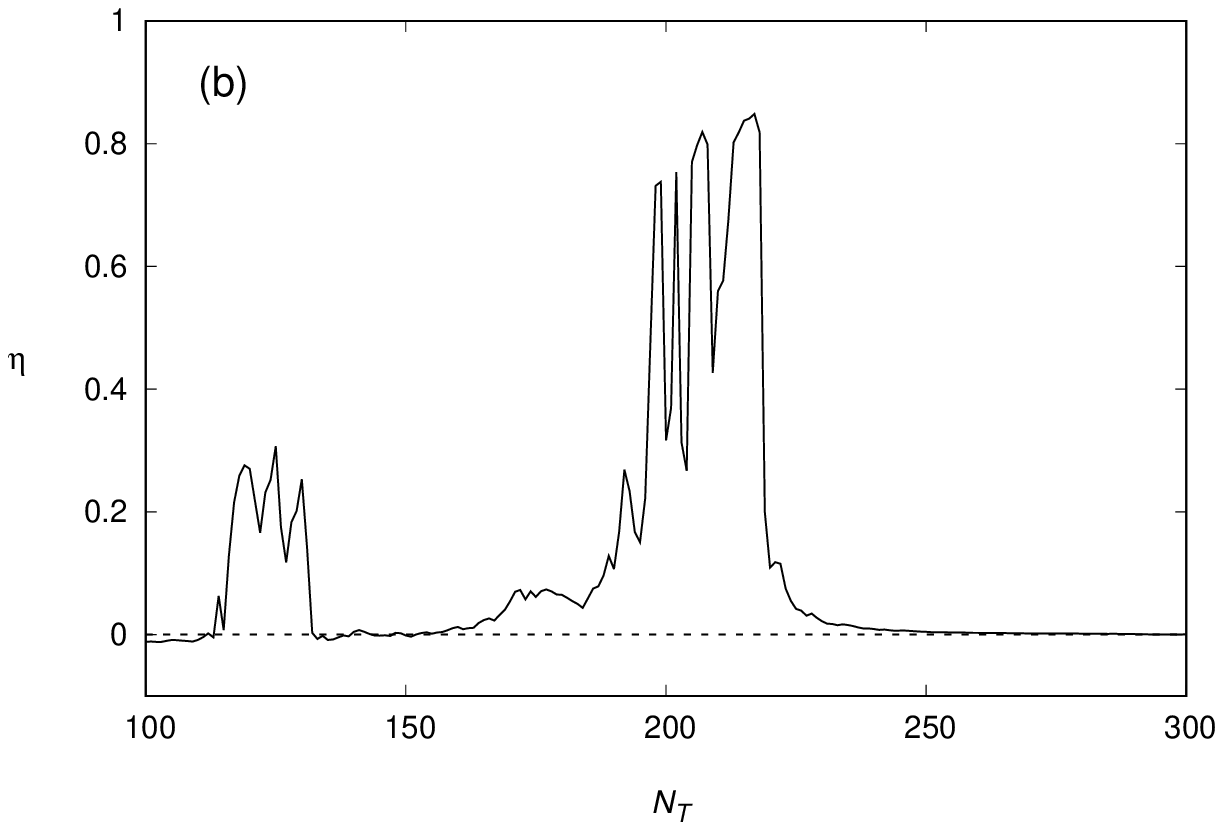}
\caption{Results for simulation shown in Fig.~\ref{simulaxy}.
(a) Inclination angles in degrees ($\alpha$) for the cluster
(solid line) and field (dashed line). (b) The corresponding
values of the parameter $\eta$ (for clarity only the region
$100 \leq N_T \leq 300$ is shown).}
\label{simulaeta}
\end{figure*}
This procedure is similar to the algorithm applied by
\citet{Gut09} to extract YSO cores using {\it Spitzer} data
\citep[see also][]{Koe08,Beu17}. They used the cumulative
distribution of branch lengths which were fitted to two
or three lines to find the transition point. Our tests
have shown that the mean branch length works better than
the cumulative length to detect the transition in certain
extreme cases, as for instance samples with low density
contrast between cluster and background.
Moreover, contrary to the above mentioned works,
we do not need to define a cut-off length to
determine the cluster radius.

\subsection{The transition parameter $\eta$}
\label{seceta}

If a cluster appears as an overdensity in the
proper-motion vector point diagram
then a cluster-field transition
point should be discernible in a $L_T-N_T$ plot,
although its exact shape and strength depend on
the data sample (see Section~\ref{secsimula}).
Our tests on both simulated and real data indicate
that in the cases of overdensities visible by eye
in the proper-motion vector point diagram
the corresponding
transition point is also clearly visible. For
the sake of a fully automatic data processing we
developed a subroutine to detect the transition point.
First, we normalize $L_T$ and $N_T$ between $0$ and
$1$ in order to make a data-independent analysis.
Second, we fit straight lines to the left and right
sides of the possible transition point. We require
a minimum of $N_{min}$ data points for the left-side
fitting to avoid noisy data effects. For the
right-side fitting we use only the first $N_{min}$
data points because in general we do not expect a
simple straight-line behaviour. From here we
calculate the inclination angles of the left-side
($\alpha_c$) and right-side ($\alpha_f$) fits.
What we do is to span all the possible transition
points ($N_T$ values) and to search for the point
where $\alpha_f - \alpha_c$ is maximum while $\alpha_c$
is minimum. The best theoretical expected transition
would be when $\alpha_f = \alpha_{max}$ ($90$~deg or
$\pi/2$~rad) and $\alpha_c = 0$~deg\footnote{This is
true for homogeneously distributed cluster stars.
In the case of distributions with steep radial
profiles $\alpha_c > 0$ (Section~\ref{secsimula}).}.
We define the dimensionless parameter
\begin{equation}
\eta=\frac{(\alpha_f-\alpha_c)}{\max\{\alpha_c,\delta\}}
\times \frac{\delta}{\alpha_{max}}
\label{eqeta}
\end{equation}
that ``quantifies" the sharpness of the transition
with a value between $0$ (no transition) and $1$
(maximal transition). The arbitrary constant $\delta$
is introduced only to prevent the singularity when
$\alpha_c=0$. Our tests on simulated data indicate
that, although the exact value of $\delta$ affect
the maximum of $\eta$ ($\eta_{max}$), it has very
little effect on the position of that maximum, that
is on the $N_T$ value at which the maximum occurs, as
long as $\delta$ is small compared with $\alpha_{max}$.
Here we are using $\delta=0.01\alpha_{max}$.
We must point out that the functional form of $\eta$
is arbitrary, but this is not really relevant as long
as we get the cluster-field transition point. The
relevance of quantifying in some way the strength
of the transition is to compare solutions obtained
with different subsamples from the same dataset
(Section~\ref{secradio}).

Fig.~\ref{simulaeta} shows the inclination angles $\alpha_c$
(solid line) and $\alpha_f$ (dashed line) and the corresponding
$\eta$ values for the well-behaved simulation shown in
Figs.~\ref{simulaxy} and \ref{simulamedia}.
For $N_T \gtrsim 300$ we see that $\alpha_c > \alpha_f$ (negative
values for $\eta$). There is a relatively narrow region around
$N_T \sim 200$ with valid solutions\footnote{We additionally
require the optimal solution to satisfy the condition
$0 \leq \alpha_c < \alpha_f$.} for which we can clearly see
that $\alpha_f - \alpha_c$ is high whereas $\alpha_c \simeq 0$.
The optimal solution (that with the maximum $\eta$ value, see
Fig.~\ref{simulaeta}b) is not located exactly at the theoretical
value $N_T = 200$ because of contamination by field stars lying
below the cluster region.

In the end, we have only one relevant free parameter: the
minimum number of data points required to get ``valid"
measurements ($N_{min}$). Its exact value is not critical
when constructing the MST because the method is almost
insensitive to the starting point. However, $N_{min}$ is
important because it determines the range of $N_T$ values
in which $\eta$ is calculated. If we have a sample of
$N_{dat}$ stars then $\eta$ can be calculated only in
the range $N_{min} < N_T < N_{dat}-N_{min}$.
This means that the algorithm will not find the optimal
solution if the actual number of cluster star is, for
instance, below or too close to $N_{min}$.
A value around $\sim \sqrt{N_{dat}}$ would be a reasonable
choice, assuming Poissonian statistics, but to be
conservative and after several tests we have assumed
$N_{min} = 3 \sqrt{N_{dat}}$.
The fact of having only one relevant free parameter
brings robustness to the algorithm because minimize
its sensitivity to parameter variations.


\subsection{Tests on simulated data}
\label{secsimula}

During the development of this algorithm we have
performed a number of tests on simulated data.
Simulations included scenarios with different
sample sizes, geometrical shapes, radial density
profiles and cluster-field density contrasts.
In general the algorithm worked quite well for
all the simulations.
The shape in which stars are distributed
(including filamentary distributions) does
not affect the detection of the cluster-field
transition point as long as the number of
member stars is larger than $N_{min}$.
Obviously,
the method works better when the surface
density of cluster stars ($\Sigma_c$) is
significantly higher than the density of
field stars ($\Sigma_f$). In fact, the
density contrast $\Sigma_c/\Sigma_f$ is
practically the only factor that determines
the behaviour and performance of the proposed
algorithm.

\begin{figure}
\centering
\includegraphics[width=0.66\columnwidth]{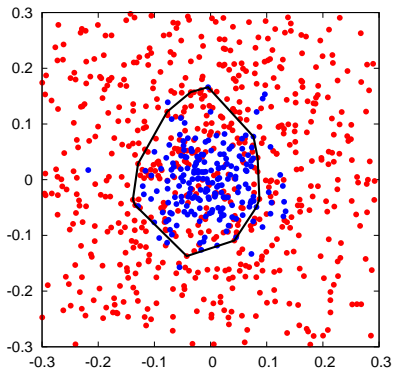}
\caption{Simulation of $200$ cluster stars (blue points) and
$800$ field stars (red points) following Gaussian density
profiles. The ratio of standard deviations between cluster
and field is $\sigma_c / \sigma_f = 0.3$. For clarity, only
the central region is shown (the whole area is circular with
radius $1$). Solid line shows the final convex hull surrounding
the selected overdensity.}
\label{simulaperfilxy}
\end{figure}

In this section we discuss some example simulations
for the case in which cluster and field follow
radial density distributions, such as is the case
for most of the real proper motion distributions.
Cluster and field stars were distributed according
to 2-dimensional Gaussian distributions having standard deviations
of $\sigma_c$ and $\sigma_f > \sigma_c$ (cluster more concentrated
than field), respectively, and both centred on the same coordinate
(this is the worst case, i.e. the most difficult to separate cluster
from field). The tests performed using elliptical (rather than
circular) distributions for the stars yielded essentially the
same results and trends. 
The only relevant variable is the
ratio $\sigma_c / \sigma_f$, strongly related to the inverse
of the density contrast $\Sigma_c / \Sigma_f$.
Fig.~\ref{simulaperfilxy} shows an example for which
$\sigma_c / \sigma_f = 0.3$,
equivalent to having an average density contrast in
the central region (within one cluster standard deviation)
of $\Sigma_c/\Sigma_f \sim 2$.
The convex hull indicates the boundary of the
overdensity according to the algorithm.
As before, some field stars fall below the overdensity area
and, additionally, some cluster stars at the edges of the
distribution (where the local cluster density is around or
below the field density at that point) are located outside
the selected overdensity.
We reiterate that this is not a
limitation of the method but consequence of how the data
sample is distributed, and can only be corrected by using
additional spatial, kinematic or photometric information.
In this work we do not intend to provide kinematic
memberships. Our aim is to determine the cluster
size in an objective and reliable manner as long
as the cluster shows an overdensity in the proper
motion space. Any cluster member we refer to is
actually a star located in the overdensity region.
Thus, at this point the algorithm selects
the ``best" boundary for a given overdensity,
that is the boundary for which the most pronounced
transition from short to long branches takes place.

\begin{figure}
\includegraphics[width=\columnwidth]{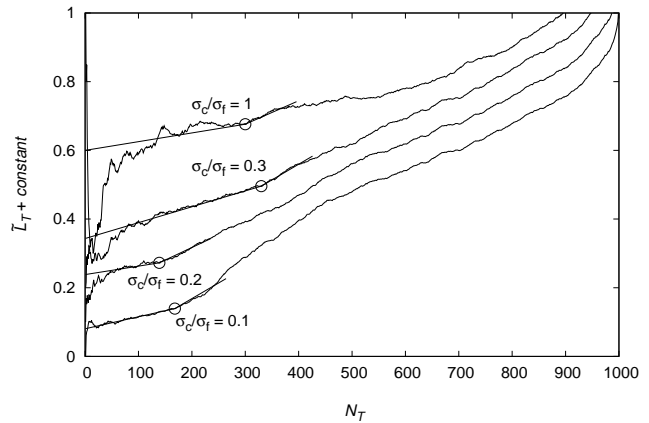}
\caption{Normalized mean length of the branches ($\tilde{L}_T$)
as a function of the number of points in the MST ($N_T$) for
four simulations with radial density distributions.
For clarity an arbitrary offset have been added to the curves.
The number of cluster and field stars are $N_c=200$ and $N_f=800$,
respectively, but the simulations are for different
cluster-to-field standard deviation ratios:
$\sigma_c/\sigma_f=0.1$, $0.2$, $0.3$ and $1$.
The corresponding average density contrasts (within a
1-$\sigma_c$ radius)
are $\Sigma_c/\Sigma_f \sim 20$, $5$, $2$ and $1$, respectively.
Open circles indicate the point where the algorithm selects the
best cluster-field transition point. The cluster and field
straight-line fits are also shown.}
\label{simulaperfilmedias}
\end{figure}

In Fig.~\ref{simulaperfilmedias} we see the $\tilde{L}_T$-$N_T$
plot for some example simulations with radial density profiles
going from a high density contrast to the no-cluster case.
Differently to the homogeneous case (Fig.~\ref{simulamedia}),
the mean length of the
branches increases as $N_T$ increases even for low $N_T$.
However, the point with a notable change in the average
slope is visible especially for the high contrast cases.
Even when $\sigma_c/\sigma_f = 0.3$ (low density contrast)
the algorithm finds
the transition point for $N_T = 331$ which corresponds quite
well to the overdensity observed in Fig.~\ref{simulaperfilxy}.

\begin{figure}
\includegraphics[width=\columnwidth]{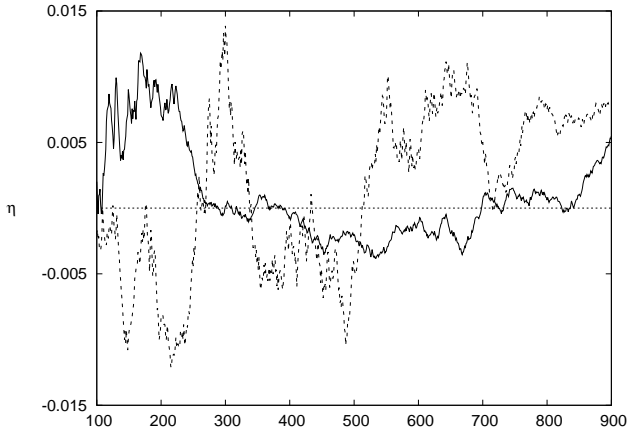}
\caption{Transition parameter $\eta$ for simulations of
radial distributions with two values of cluster-to-field
standard deviation ratios: $\sigma_c/\sigma_f=0.1$ (solid
line) and $\sigma_c/\sigma_f=1$ (dashed line).}
\label{simulaperfileseta}
\end{figure}

The simulation labelled $\sigma_c/\sigma_f=1$ corresponds
to the case when there is no cluster but just one radial
distribution of stars. This case exhibits random fluctuations
from which the algorithm simply selects the strongest change
in slope. This can be seen in Fig.~\ref{simulaperfileseta}
that shows $\eta$ for the high density contrast and
no-cluster cases.
Although the values of $\eta$ at high contrast
are relatively low compared with the well-behaved,
homogeneous distribution (see Fig.~\ref{simulaeta}b),
its maximum value clearly stands out at $N_T=169$.
It is different for the no-cluster case (dashed line in
Fig.~\ref{simulaperfileseta}) where the selected optimal
transition at $N_T=301$ is not very different from other
local maxima of $\eta$ around $N_T \sim 600$.

\subsection{Estimation of cluster radius}
\label{secradio}

It is not a simple task to determine (to define) the cluster
radius in an objective way because the definition of radius
is ambiguous itself given the great variety of observed
morphologies. There are several commonly used characteristic
radii, such as the core radius, half-mass (or half-light)
radius, tidal radius, or simply the ``extent" of the cluster
usually defined as the radius where the cluster surface
density drops below field density (the details depend on
the author). Mixing these different concepts can lead to
inaccurate or biased global results
\citep[see discussion in][]{Pfa16}.
Here we use the simple geometric definition of cluster radius
as the radius of the smallest circle containing all the cluster
stars. Following graph theory terminology we can refer to it
as the covering radius ($R_{cov}$) to differentiate it of
other characteristic radii. According to \citet{San10} the
sampling radius $R_s$ (i.e. the radius of the circular
area around the cluster position used to extract the data
from a given catalogue) that best discriminates kinematic
members from field stars is $R_s=R_{cov}$. In this case
an overdensity corresponding to the cluster's centroid should
be visible in the proper motion space. The procedure explained
in the previous sections determines in a simple and direct way 
the area covered by this overdensity. For $R_s < R_{cov}$
the cluster is subsampled (by varying amounts, depending on
cluster star density). On the contrary, for $R_s > R_{cov}$
only new field stars are included so that the cluster overdensity
will be less prominent. The transition parameter defined in
Section~\ref{seceta} quantitatively measures the sharpness
of the overdensity. Thus, the strategy we follow is to
apply an external loop over a range of $R_s$ values
and, at each step calculate the maximum transition
parameter $\eta_{max}$. We consider the optimal sampling
radius as that with the highest of all the $\eta_{max}$
values. This optimal sampling radius is, as already
discussed, the most reliable estimation of the actual
cluster covering radius. Thus, this approach give us a
method to calculate cluster radii directly from the data
without making any additional assumptions about the spatial
distribution of the cluster stars.

\subsection{Estimation of uncertainties}
\label{secerrores}

The value of $\eta$ that determines the boundary of the
overdensity ($\eta_{max}$) is unique for each given sampling
radius $R_s$. We have estimated an uncertainty associated
to each $\eta_{max}$ value using bootstrap techniques: we
repeat the calculation on a series of random resamplings
of the data, and the standard deviation of the obtained 
set of $\eta_{max}$-values is taken as the error in our
estimation.
Additionally, we use this error as a reference to
estimate an overall uncertainty associated with the
derived cluster radius. For this we define a lower
limit given by the optimal solution (the highest
of all the $\eta_{max}$ values) minus three times
its standard deviation, and we assume that the range
of acceptable solutions for the radius are all the
values for which $\eta_{max}$ is above this lower
limit (see Fig.~\ref{fig188} in next section for
an example).

\subsection{Test on NGC~188}
\label{secngc188}

The direct way of validating our method is to apply it
to a well-known open cluster and compare the results.
NGC~188 serves as a test case because it is old (and
therefore it exhibits a clear radial density profile)
and it is located far above the Galactic plane (with
little contamination by field stars). NGC~188 has
been extensively studied and it has relatively
well-determined physical parameters
\citep[see Table~1 in][for a summary of some
published parameters]{Els16}.
Regarding the cluster size, the radius reported
in \citet{Dia02} for NGC~188 is 8.5~arcmin, which 
is the value given in the WEBDA database \citep{Mer95},
whereas \citet{Kha13} estimated a relatively high value
of 34.2~arcmin. \citet{Sam17} determined a radius of
12~arcmin from its radial density profile.
The characteristic scale most similar to what we
call the covering radius is the {\it limiting}
radius $R_{lim}$ defined as the radius that covers
the cluster and reaches ``enough" 
stability\footnote{This usually means
that the surface star density equals the background 
density plus three standard deviations.}
with the background \citep{Tad14}.
\citet{Bon05} estimated $R_{lim}=24\pm0.1$~arcmin
for NGC~188, almost twice the last value of
$R_{lim}=12.45$~arcmin given by \citet{Els16}.
These dissimilar values serve to exemplify the
necessity of alternative approaches such as the
one proposed here.

\begin{figure}
\includegraphics[width=\columnwidth]{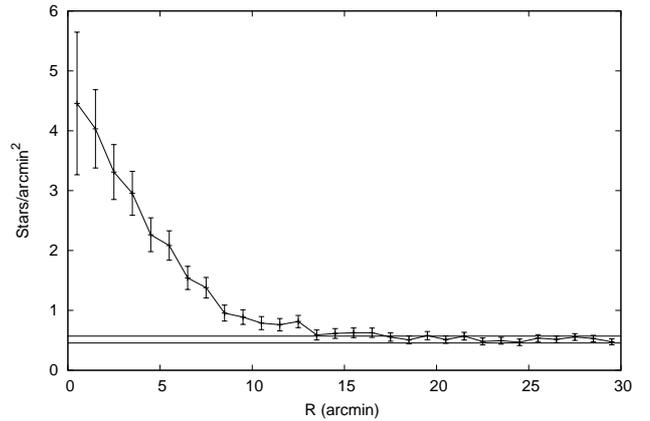}
\caption{
Radial density profile of stars toward NGC~188.
Error bars are from Poisson statistics. Horizontal dashed
line indicates the mean value of $0.51$ stars/arcmin$^{2}$
(plus/minus one standard deviation) estimated beyond
$20$~arcmin.}
\label{figperfil}
\end{figure}

In order to test the method with NGC~188 we extract
its data (positions and proper motions) from the UCAC4
catalogue \citep{Zac13} in the same way that we will do
with the rest of the cluster (Section~\ref{seccumulos}).
Figure~\ref{figperfil} shows the corresponding radial
density profile. Clearly the cluster density profile
merges into the background at some point around
$\sim 15$~arcmin (the exact value depending on the
specific merging criterion).

\begin{figure}
\includegraphics[width=\columnwidth]{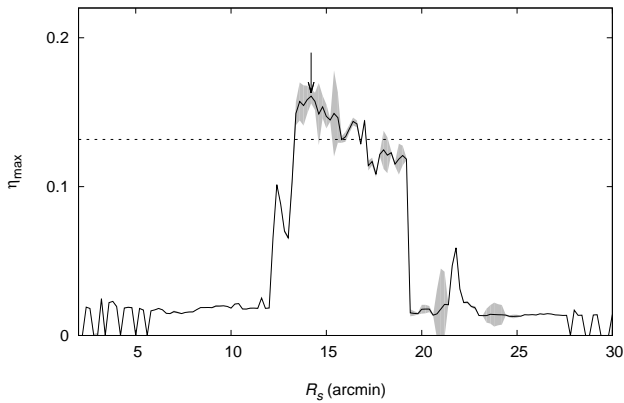}
\caption{
Maximum value of the transition parameter ($\eta_{max}$)
as a function of the sampling radius ($R_s$) for the open
cluster NGC~188 (solid black line) with the vertical arrow
indicating the obtained maximum. Grey shadow corresponds
to one standard deviation computed using bootstrapping,
whereas the horizontal dashed line indicates the maximum
$\eta_{max}$ minus three standard deviations.}
\label{fig188}
\end{figure}

The result of applying our algorithm to NGC~188 is
shown in Figure~\ref{fig188}. For this well-behaved
open cluster the maximum value of $\eta_{max}=0.16$
is found at $R_s=14.2$~arcmin, in very good
agreement with its spatial density profile in
Figure~\ref{figperfil}.
As explained in Section~\ref{secerrores}, we associate 
an uncertainty to the calculated radius by considering 
all the $\eta_{max}$ that are above the dashed line in
Figure~\ref{fig188}. In this case the final cluster
radio would be in the range $R_c = 13.4-17.0$~arcmin.

In next sections we will apply this same procedure 
to a sample of open cluster with discrepant radius 
values in the literature.

\section{Sample of clusters}
\label{secsample}

We have compared the open cluster catalogues of \citet{Dia02}
and \citet{Kha13} (hereafter D02 and K13, respectively), both
available via the VizieR\footnote{http://vizier.u-strasbg.fr}
database. The latest version (3.5 of 2016 February) of D02
contains updated information on $2167$ optically visible open
clusters and candidates. \citet{Dia14} used the UCAC4 catalogue
\citep{Zac13} to determine in a homogeneous way kinematic
memberships and mean proper motions for most of these clusters.
However, the apparent radii of many of the clusters in D02 were
compiled from older references \citep[e.g.][]{Lyn87,Mer95} in
which most of the apparent diameters were estimated from visual
inspection. On the other hand, K13 used data from the PPMXL
\citep{Roe10} and 2MASS \citep{Skr06} catalogues to calculate
and provide a set of uniform astrophysical parameters for
$3006$ clusters (most of them open clusters). They used
multi-dimensional diagrams to determine combined (kinematic
and photometric) membership probabilities for the stars
\citep{Kha12}. They calculated cluster sizes fitting by eye
the radial density profiles of the 1-$\sigma$ members. The
fitting uses three empirical parameters: the radius of the
core, of the central part and of the cluster. We take the
last one, defined in K13 as the distance from the cluster
centre where the surface density of members becomes equal
to the average density of the field, as their estimation
for the cluster radius.
Cluster radius distributions for the full D02 and K13
catalogues exhibit an apparent bias with systematically higher
cluster radii in K13 (that we will denote as $R_K$) than in
D02 ($R_D$).
The mean $R_D$ value is $\sim 7.2$~arcmin 
and the median $\sim 2.5$~arcmin, whereas
the mean of $R_K$ is $\sim 10.5$~arcmin and the
median $\sim 7.8$~arcmin.

\begin{figure}
\includegraphics[width=\columnwidth]{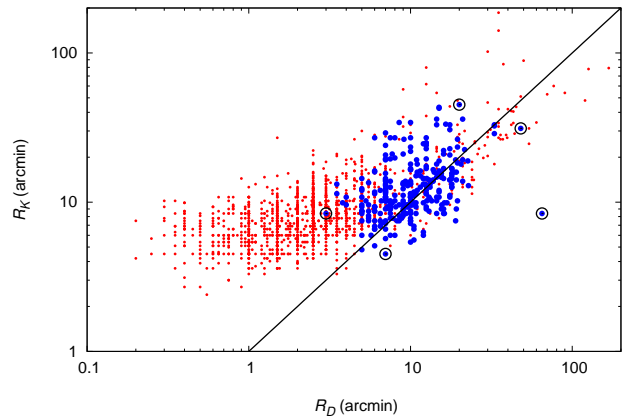}
\caption{Cluster radius in the K13 catalogue ($R_K$) as a
function of radius in D02 ($R_D$). All the clusters in
common between both catalogues are shown as red points,
cluster having more than $400$ expected members according
to D02 are shown as blue points, and selected clusters are
indicated with surrounding open circles. The solid line
indicates the 1:1 line.}
\label{figradios}
\end{figure}

In order to select a sample of clusters that were certainly
common to both catalogues, we first matched each cluster in
D02 with the closest cluster in K13, and then we kept only
those with the same main name in both catalogues. These
last step may accidentally discard some common clusters
but in this way we are pretty sure we are comparing the
same clusters. The matching yields $1706$ clusters whose
radii are plotted in Fig.~\ref{figradios}.
Again we can
see that $R_K$ tends to be systematically higher than $R_D$,
especially for small radius values ($R_D < 3$~arcmin).
We need a sufficiently high number of cluster stars to reach
a valid solution ($N_c$ has to be greater than $N_{min}$).
We do not know in advance $N_c$ but D02 give an estimation
of the expected number of cluster members. They warn this
can be an overestimate (our tests showed us that their
estimations use to be $3-4$ times our final $N_c$ value),
thus we additionally require the number of members estimated
in D02 to be higher than $400$. The resulting $284$ clusters
are shown as blue dots in Fig.~\ref{figradios}.
From here we select five clusters having extreme radii in 
the D02 and K13 catalogues: Ruprecht~175 (with the smallest
$R_K$ value), NGC~6603 (the smallest $R_D$), NGC~1647 (the 
largest $R_K$), and ASCC~19 and Collinder~471 (the two largest
$R_D$ values). A comparison of the cluster radii reported in
D02 and K13, and the results obtained in this work is shown
in Table~\ref{tabcumulos}.
\begin{table*}
\centering
\caption{Properties of the selected clusters.}
\label{tabcumulos}
\begin{tabular}{lcccccc}
\hline
     &       &       & \multicolumn{3}{c}{Cluster radius} \\
\cline{4-6}
Name & RA      & DEC                       & D02      & K13      & This work \\
     & (h m s) & ($\degr~\arcmin~\arcsec$) & (arcmin) & (arcmin) & (arcmin) \\
\hline
NGC~188       & $00~47~28$ & $+85~15~18$ &  8.5 &  34.2 & 15.2 $\pm$ 1.8 \\
NGC~1647      & $04~45~55$ & $+19~06~54$ & 20.0 &  45.0 & 29.4 $\pm$ 3.4 \\
ASCC~19       & $05~27~47$ & $-01~58~48$ & 48.0 &  31.2 & ...            \\
NGC~6603      & $18~18~26$ & $-18~24~24$ &  3.0 &   8.4 &  4.2 $\pm$ 1.7 \\
Ruprecht~175  & $20~45~12$ & $+35~30~00$ &  7.0 &   4.5 &  7.0 $\pm$ 0.3 \\
Collinder~471 & $22~07~06$ & $+72~00~00$ & 65.0 &   8.4 & ...            \\
\hline
\end{tabular}
\end{table*}

\section{Radius determination for the selected clusters}
\label{seccumulos}

We use data from the UCAC4 catalogue \citep{Zac13} to apply the
proposed method to the open clusters listed in Table~\ref{tabcumulos}.
We extract proper motions for all the stars and clean the data leaving
only ``good" stars\footnote{We excluded double systems and stars with
known problems (overexposed, high proper motion, poor astrometric
solution), i.e. we required the UCAC4 flags db=0 and of=0.}.
We run the program spanning a wide range of sampling radius ($R_s$)
values including both $R_D$ and $R_K$, with steps of $0.1-0.2$~arcmin.

\subsection{NGC~1647}

\begin{figure}
\includegraphics[width=\columnwidth]{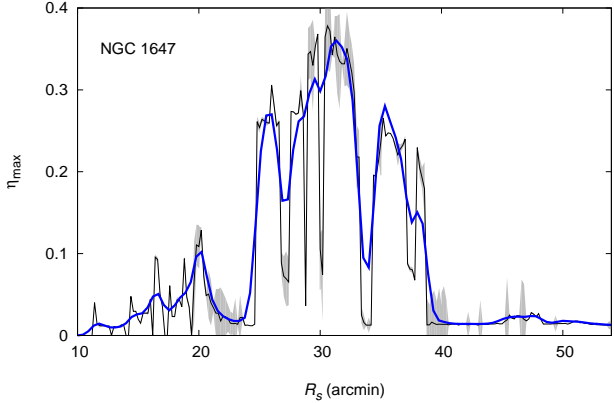}
\caption{Maximum value of the transition parameter ($\eta_{max}$)
as a function of the sampling radius ($R_s$) for the open cluster
NGC~1647. The superimposed blue line is a smoothed curve obtained
by using a Gaussian kernel (see text).}
\label{figsamplingA}
\end{figure}

We will discuss in detail the first cluster of the 
selected sample
(NGC~1647). Fig.~\ref{figsamplingA} shows the obtained
$\eta_{max}$ value for each sampling radius $R_s$.
The first thing we note is that $\eta_{max}$ fluctuates with
small variations in $R_s$. This is due to the functional
form of the transition parameter (Eq.~\ref{eqeta}). The 
maximum $\eta$ values tend to occur for small values of
$\alpha_c$ so that small $\alpha_c$ variations imply
noticeable variations of $\eta$ (as can be seen in
Figure~\ref{simulaeta}).
This effect will be more or less noticeable depending
on the data itself and on how clear the overdensity
can be seen in the proper motion space; for instance
in NGC~188 these fluctuations are less apparent
(Figure~\ref{fig188}).
Despite this, the overall trend is clearly discernible
for NGC~1647,
with some local maxima and an absolute maximum around
$R_s \simeq 30$~arcmin. For a better visualization
of the global behaviour we have superimposed a smoothed
function (blue line in Fig.~\ref{figsamplingA}). The
smoothing is done by convolving the data with a Gaussian
kernel. According to \citet{Sil86}, the optimal bandwidth
for $N$ normally distributed data points with standard
deviation $\sigma$ is around $\sigma(4/3N)^{0.2}$,
although this usually yields very conservative broad
bandwidth so that we always use $1/10$ of the
Silverman's rule for the bandwidth. The most relevant
transition cluster-field for NGC~1647 corresponds to
$\eta_{max}=0.38$ 
at an optimal sampling radius of $R_s=30.6$~arcmin.
Taking the uncertainty in $\eta_{max}$ into account
our final estimation of the cluster radius is
$R_c=26.0-32.8$~arcmin.
The obtained radius is intermediate between the values
given in D02 ($20$~arcmin) and K13
$R_K=45$~arcmin, and it agrees with the $30$~arcmin estimated
by \citet{Gef96} from visual inspection of Palomar plates.
In any case, beyond the associated uncertainties,
Fig.~\ref{figsamplingA} clearly rules out large values
($R_C \gtrsim 40$~arcmin) reported in other works
\citep{Pis07,Pis08,San09,Kha13}.

\begin{figure}
\includegraphics[width=\columnwidth]{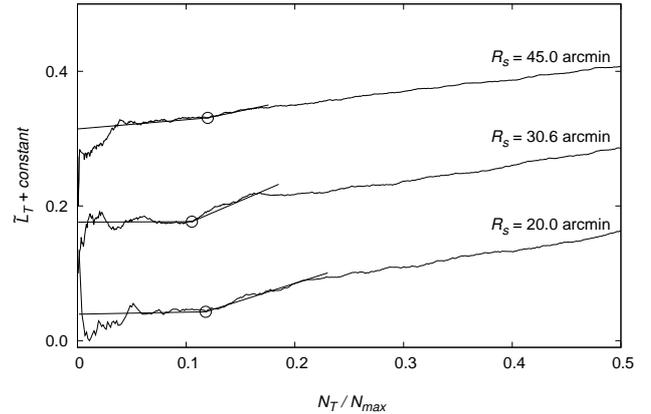}
\caption{Normalized mean length of the branches $\tilde{L}_T$
as a function of the fraction of stars in the MST $N_T/N_{max}$
for the three indicated sampling radii. An arbitrary offset
have been added to the curves. Open circles and solid lines
indicated the best choices and the fits for the transition
points.}
\label{figcumulosmedias}
\end{figure}

The particular results for three different sampling radii
are compared in Fig.~\ref{figcumulosmedias}.
\begin{figure*}
\includegraphics[width=0.66\columnwidth]{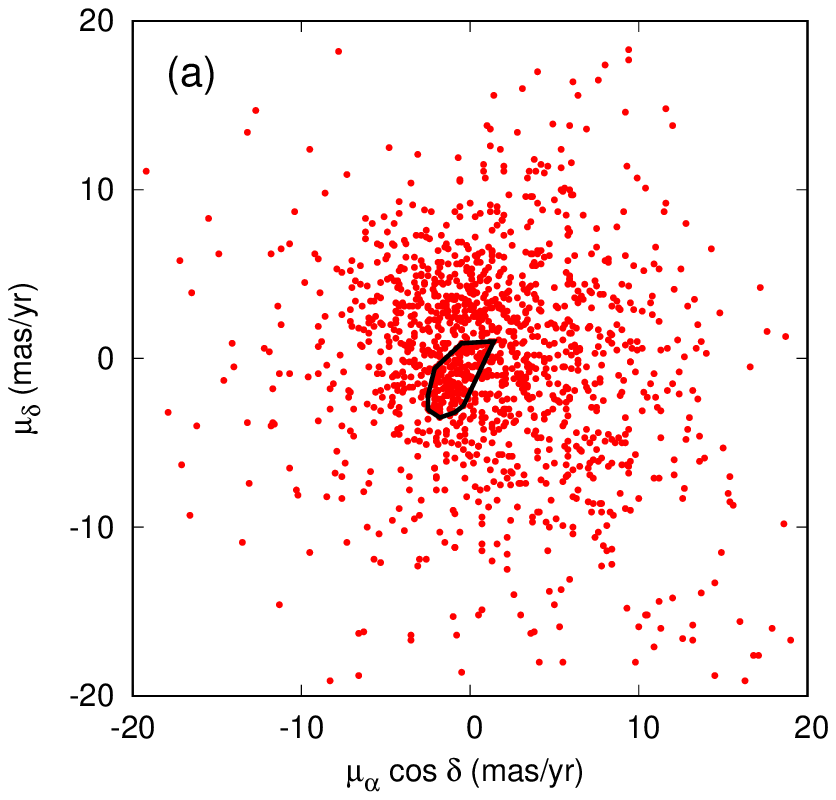}
\includegraphics[width=0.66\columnwidth]{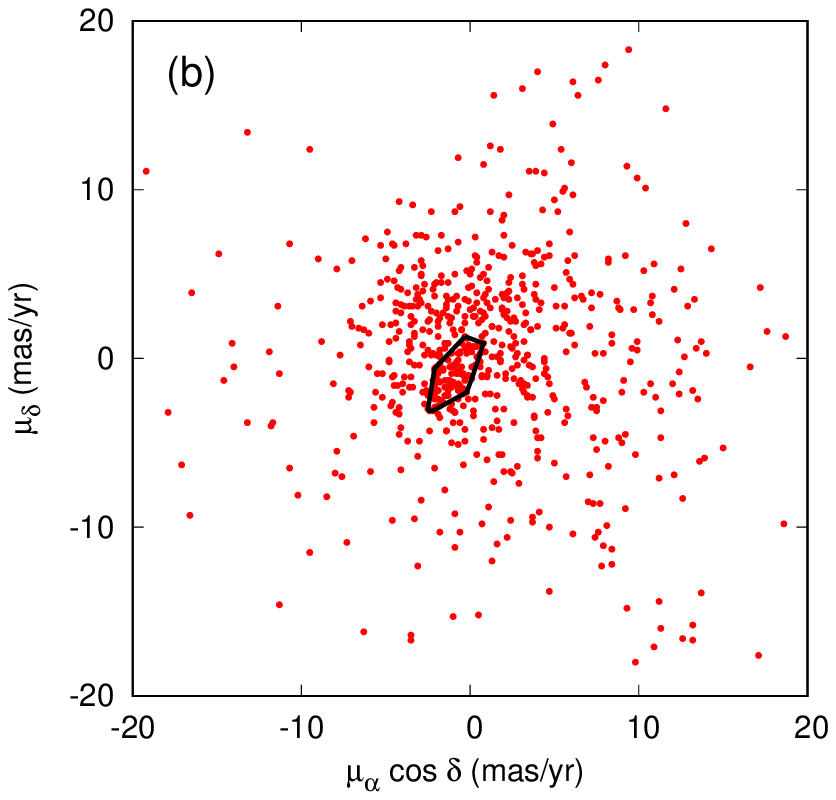}
\includegraphics[width=0.66\columnwidth]{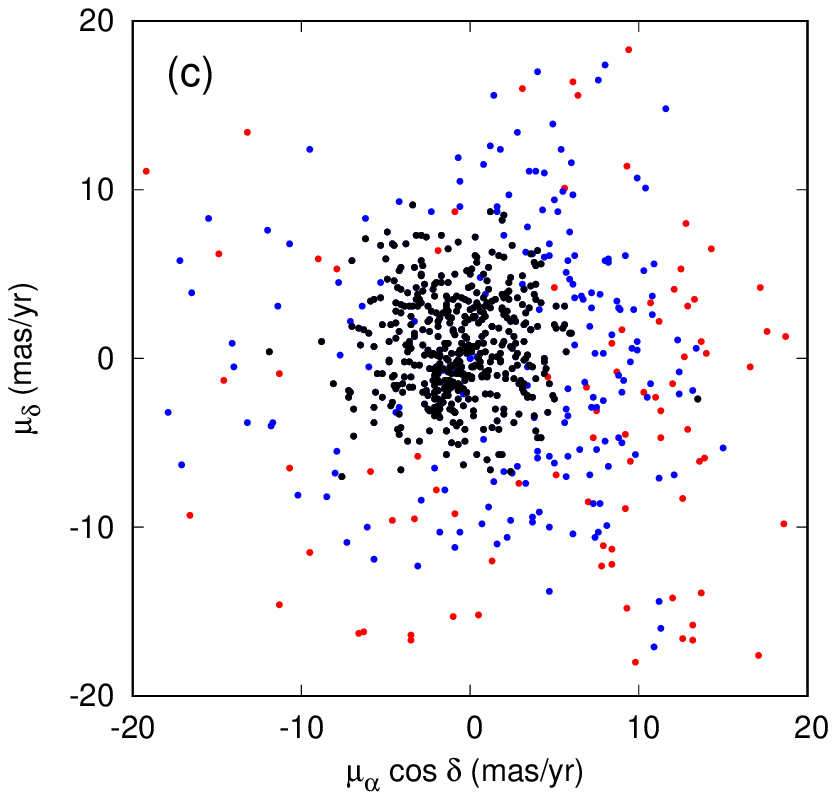}
\caption{Distribution of UCAC4 proper motions for the
stars in the field of NGC~1647. (a) All the stars (red
points) corresponding to the optimal sampling radius
$R_s=30.6$~arcmin. Solid line is the convex hull
surrounding the selected overdensity.
(b) As in (a) but for $R_s=R_D=20$~arcmin.
(c) Cluster members according to \citet{Dia14} (they
used $R_s=21$~arcmin). Blue points are star having
membership probabilities higher than $50\%$ and black
points have probabilities higher than $90\%$.}
\label{figcumMP}
\end{figure*}
For a sampling radius of $R_s=R_D=20$~arcmin 
(the value given by D02)
the sample consists of $722$ stars,
from which the algorithm selects $N_T=86$
stars inside the overdensity area
with a transition parameter of $\eta_{max}=0.15$. Instead,
when $R_s=R_K=45$~arcmin (K13) the full sample is $2897$ with
$N_T=348$ but, in this case, this is a
relatively bad solution with $\eta_{max}=0.01$ as it
can be easily seen by eye in Fig.~\ref{figcumulosmedias} (an
almost imperceptible transition for $R_s=45$~arcmin). For the
optimal sampling radius ($30.6$~arcmin, this work)
we obtain $N_T=155$
stars in the overdensity
out of a total
of $1464$ stars in the sample with a clearly detected transition
($\eta_{max}=0.38$).
It is interesting to note that the fraction of
overdensity stars
is always around $N_T/N_{max} \sim 0.1$. This fact is an
indicator of robustness of the method: the algorithm finds the
area occupied by the overdensity, and when the sampling radius
increases the number of contaminant stars also increases but
the overdensity area remains nearly constant (see below) and
$N_T/N_{max}$ changes very little.
Fig.~\ref{figcumMP} shows proper motion distributions for two
sampling radii: the optimal value found in this work (panel a)
and that corresponding to the radius $R_D$ reported in D02
(panel b).
By comparing panels a-b we
see that the selected overdensity area is nearly the same
even though the sampling radii are very different (the
number of sample stars in panel a is twice that of panel b).
This is not the case for the widely used method of fitting
two Gaussian functions to represent the distributions of
field and cluster stars \citep{Vas58,San71,Cab85}.
In this case, when the sample is contaminated by many field
stars the fit tends to produce a wider and flatter function
for the field distribution and, as a consequence, the
membership probabilities (defined as the ratio cluster-total
distributions) increase and the number of spurious members also
increases \citep[this effect has been discussed in][]{San10}.
Panels b-c of Fig.~\ref{figcumMP} compare our results with
those of \citet{Dia14} (they used proper motions from UCAC4
to fit two elliptical bivariate Gaussian functions).
As mentioned before our algorithm does not provide
cluster memberships because this MST-based procedure
only selects the area comprising the overdensity,
although obviously the $155$ stars inside the convex
hull are probable kinematic members of the cluster.
There should be other additional members beyond the
overdensity area where the cluster star density is
around or below the local field star density.
However, it is interesting to note that the number
of overdensity stars is considerably smaller than
the $459$ very probable members (membership
probabilities $\geq 90\%$) according to \citet{Dia14}
(black points in Fig.~\ref{figcumMP}) or than the
$618$ 1-$\sigma$ members found by K13.

An excessively large number of spurious members can lead
to inaccurate or biased estimations of open cluster proper
motions (and other properties).
\citet{Kur16} used both kinematic and photometric
criteria to select the most reliable members and recalculated
proper motions for a sample of $15$ open clusters. For
some of the clusters their results differ significantly
from the ones given by \citet{Dia14}, and they suggested
that the difference could be linked to a field star
contamination effect.
In the case of NGC~1647, \citet{Kur16} calculated a proper motion 
$(\mu_\alpha\cos\delta , \mu_\delta) = (-1.13,-1.27)$~mas~yr$^{-1}$
whereas \citet{Dia14} obtained
$(\mu_\alpha\cos\delta , \mu_\delta) = (-0.74,-0.57)$~mas~yr$^{-1}$.
Our cluster proper motion centroid
$(\mu_\alpha\cos\delta , \mu_\delta) = (-0.85,-1.11)$~mas~yr$^{-1}$
is in between both values but slightly closer to the 
\citet{Kur16} result ($| \Delta \mu | = 0.32$~mas~yr$^{-1}$).
However, the error in proper motions ($\sim 1-4$~mas~yr$^{-1}$)
are similar to the errors given by \citet{Kur16} and \citet{Dia14}
(limited by UCAC4 proper motion errors), so that theses differences
are not significant.

\subsection{The rest of the open clusters}

\begin{figure*}
\includegraphics[width=\columnwidth]{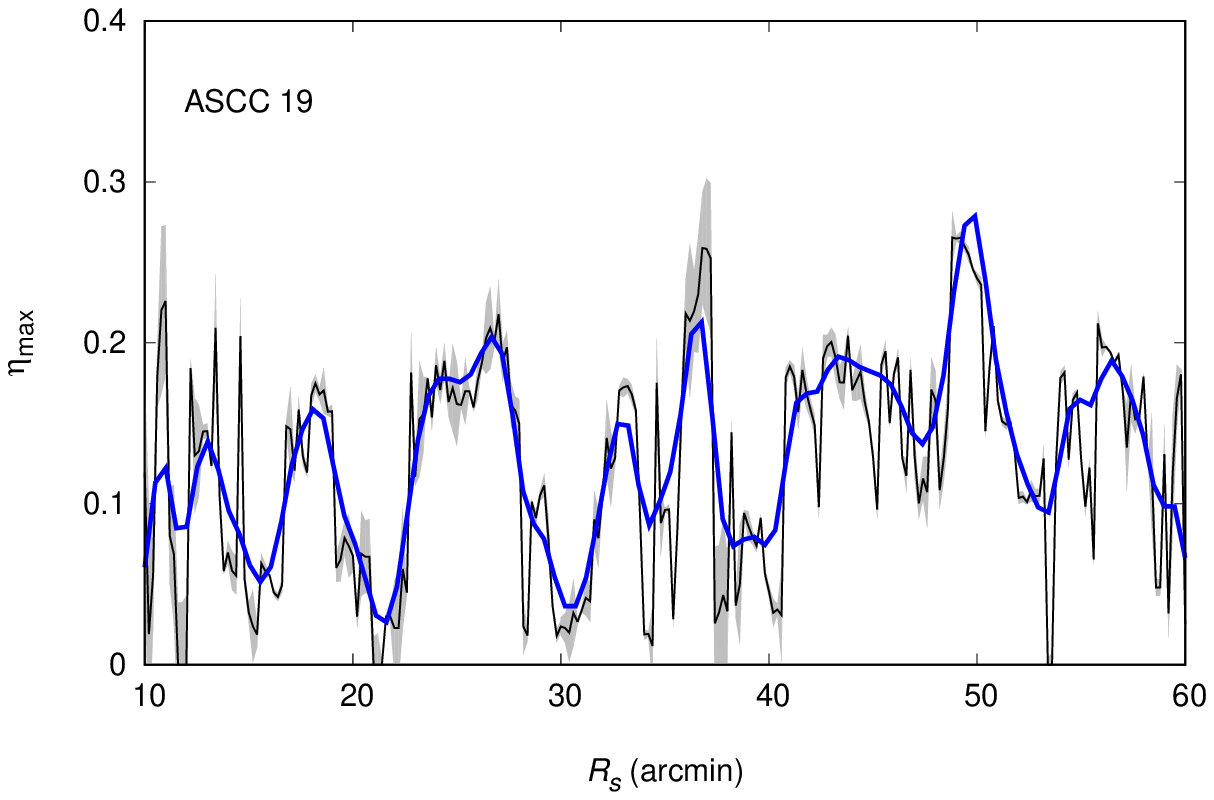}
\includegraphics[width=\columnwidth]{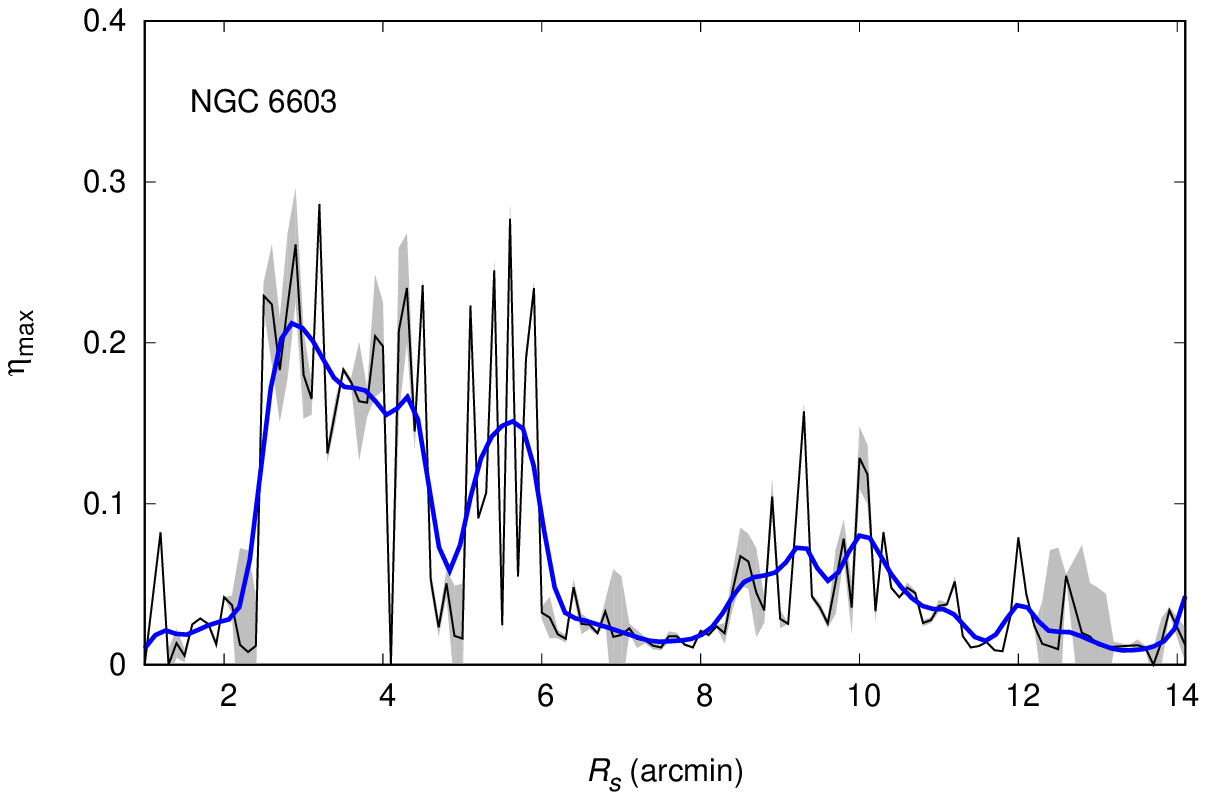}
\includegraphics[width=\columnwidth]{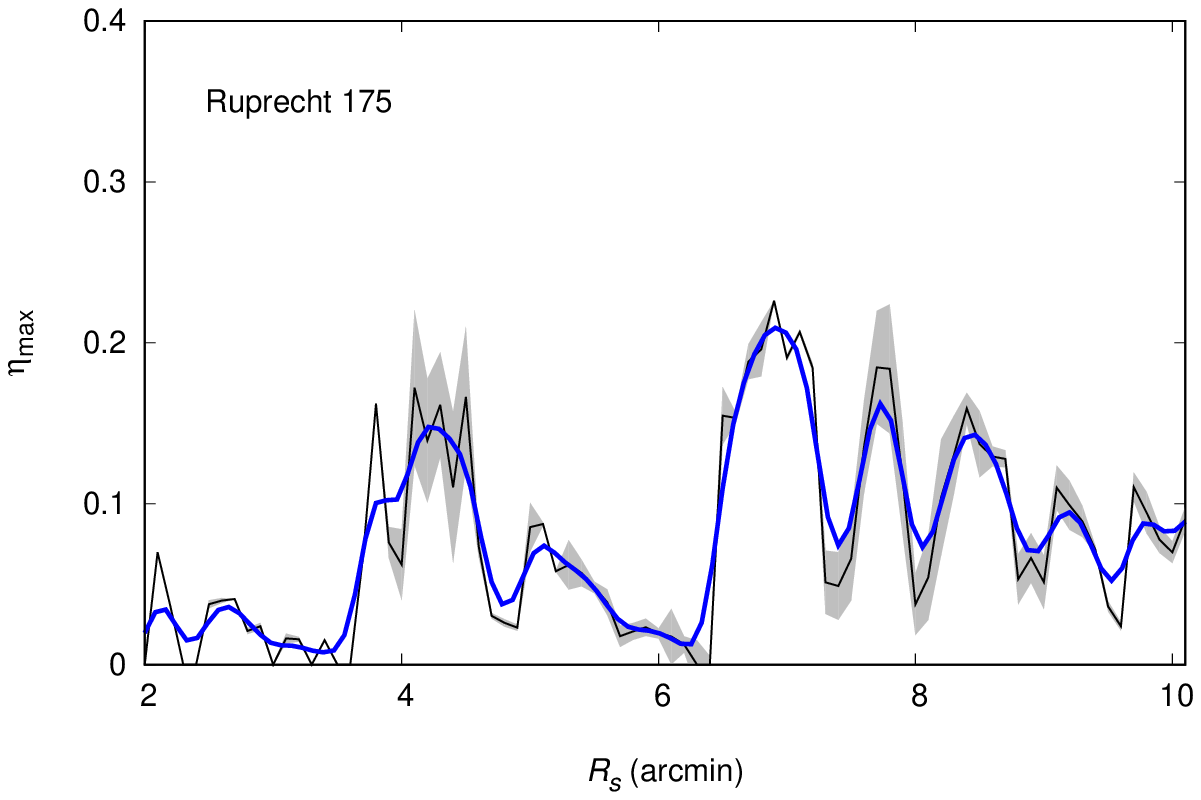}
\includegraphics[width=\columnwidth]{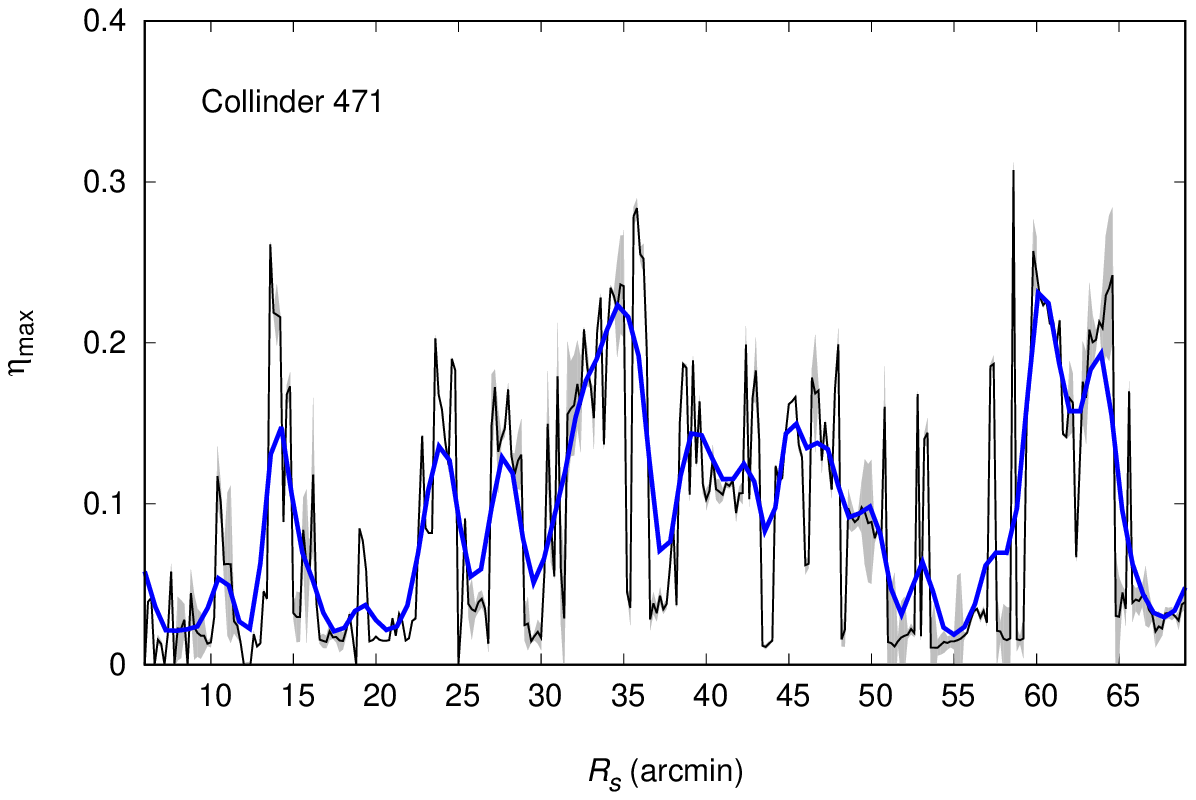}
\caption{As in Fig.~\ref{figsamplingA} but for the rest
of the open clusters listed in Table~\ref{tabcumulos}.}
\label{figsamplingBCDE}
\end{figure*}

The results for the remaining four clusters of
Table~\ref{tabcumulos} are shown in Fig.~\ref{figsamplingBCDE}.
We again see the same kind of fluctuations in $\eta_{max}$
as in Fig.~\ref{figsamplingA}. The smoothed (blue) curves
allow to focus on global trends that we will comment on
below.

\subsubsection*{\bf ASCC~19:}

\begin{figure*}
\includegraphics[width=0.66\columnwidth]{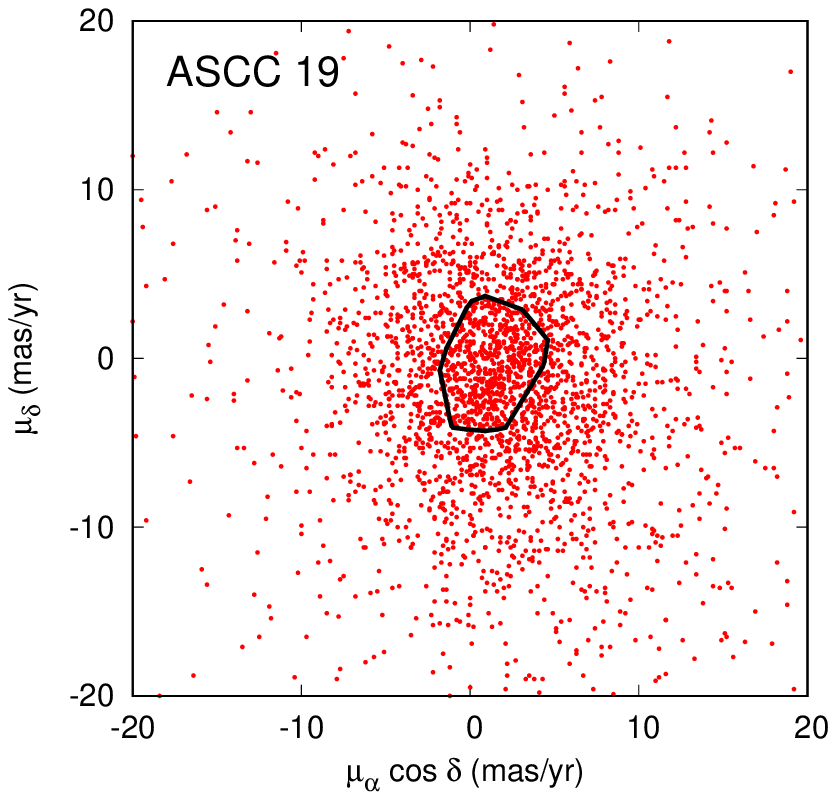}
\includegraphics[width=0.66\columnwidth]{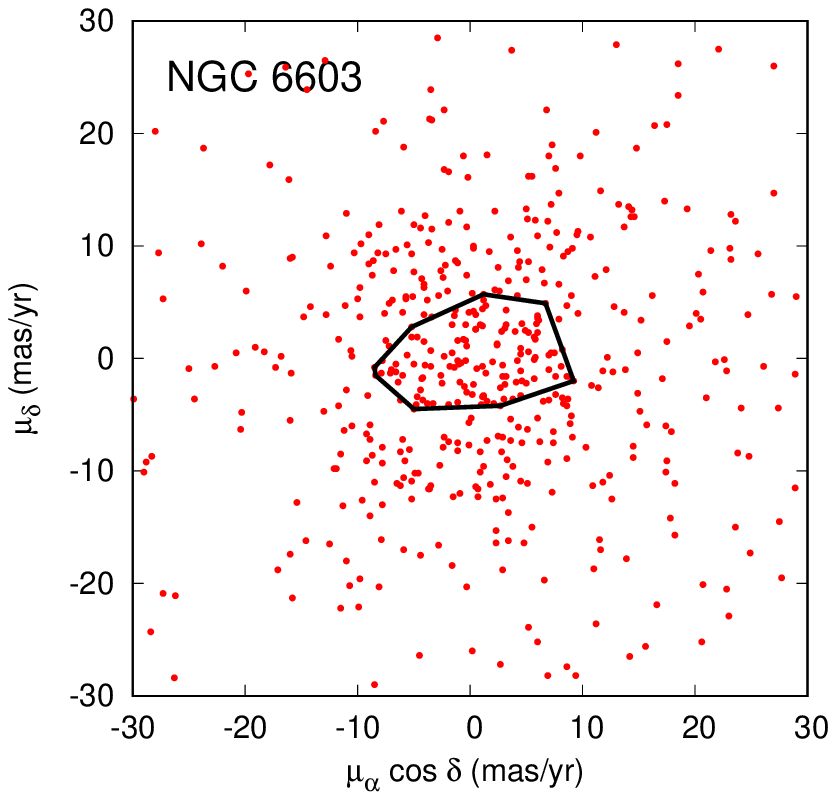}\\
\includegraphics[width=0.66\columnwidth]{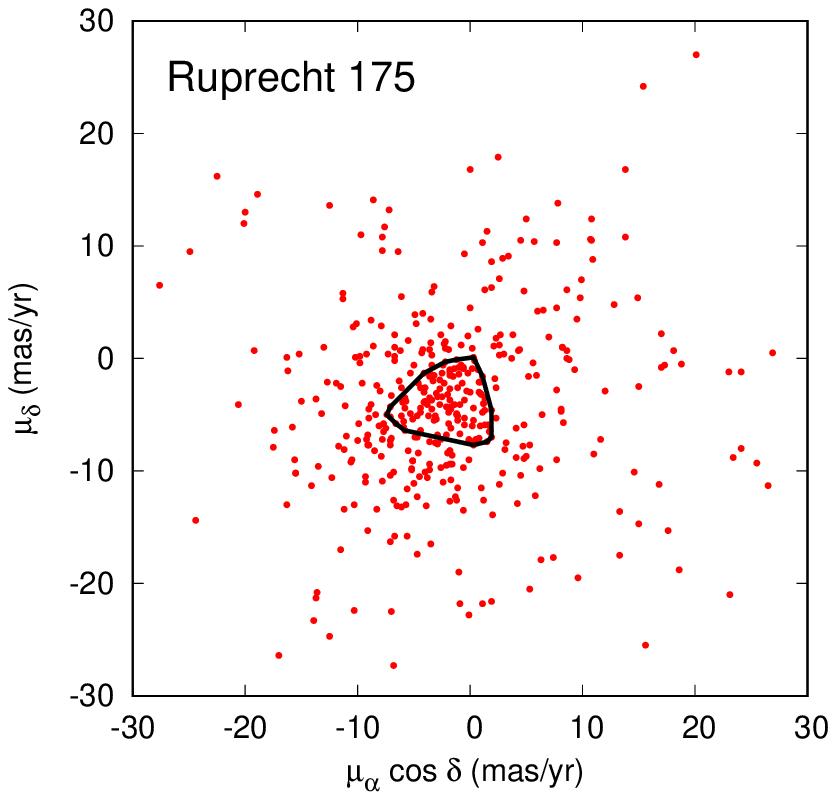}
\includegraphics[width=0.66\columnwidth]{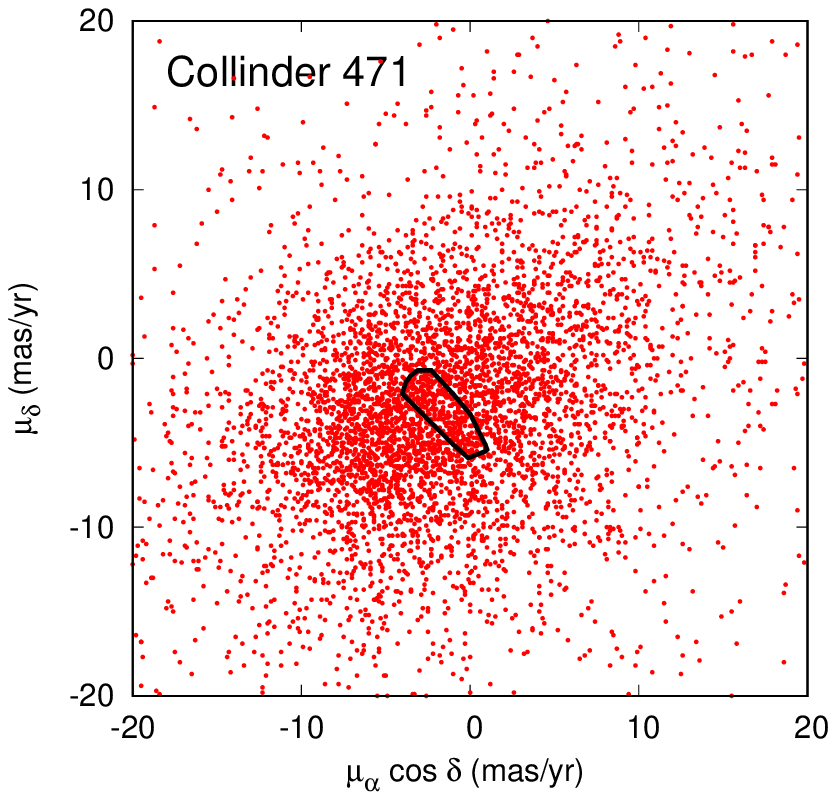}
\caption{Proper motion distributions for the rest
of the open clusters listed in Table~\ref{tabcumulos}
and for their optimal sampling radii (see text).
Solid lines are the convex hulls surrounding the
cluster members.}
\label{figcumMPresto}
\end{figure*}

This is a cluster reported as new by \citet{Kha05b} with
a radius of $48$~arcmin \citep[see also][]{Pis07} that
it is the value given in D02. Afterwards, \citet{Kha13}
recalculated a cluster radius of $31.2$~arcmin.
We spanned a wide range of $R_s$ values but it has
been not possible to find out a clear maximum for
$\eta_{max}$. 
There are several local maxima with one of them
slightly standing out at $\sim 49$~arcmin, a
value very close to that in D02.
We would like to point out that this unsuccessful
outcome does not represent a ``failure" of the method.
For a given $R_s$ the algorithm recovers the overdensity
in proper motions and the boundary that best delimits
the cluster-field transition. The problem is that
different sampling radii yield similar changes in
slope. Thus, by using only kinematic data, we are
not able to say what is the optimal sampling radius
and, therefore, the cluster radius. This may be due,
among others things, to the existence a more complex
underlying patterns or simply to the lack of a clear
overdensity in the proper motions space. An additional
analysis including other physical variables (positions,
photometry) should clarify this issue.
We prefer to be conservative and say we did not find a
feasible solution for ASCC~19.
We use $R_s=49$~arcmin to show the proper motion
distribution in Fig.~\ref{figcumMPresto}.
The number of data points for this radius is $2908$ from
which $614$ corresponds to 
stars inside the convex hull.
The proper motion centroid is at
$(\mu_\alpha\cos\delta , \mu_\delta) = (+1.20,-0.41)$~mas~yr$^{-1}$
and it is not significantly different
($| \Delta \mu | = 0.63$~mas~yr$^{-1}$)
from the value
$(\mu_\alpha\cos\delta , \mu_\delta) = (+1.43,-1.00)$~mas~yr$^{-1}$
given by \citet{Dia14}.

\subsubsection*{\bf NGC~6603:}

For this cluster, the radius reported in the literature
ranges from $R_c=2.8-3.0$~arcmin \citep{Sag98,Dia02} to
$R_c=7.2-8.4$~arcmin \citep{Kha05a,Kha13}.
Our result (Fig.~\ref{figsamplingBCDE}) points
to some value in the range $R_c = 2.5-5.9$~arcmin.
There is another local maximum
at around $\sim 9-10$~arcmin
\citep[very close to the value given by][]{Kha13},
but the former value is clearly the best solution.
For a sampling radius at the center of the obtained
range ($R_s=4.2$~arcmin)
there are $595$ stars in the sample from which $123$
are part of the overdensity, whose calculated centroid is
$(\mu_\alpha\cos\delta , \mu_\delta) = (+0.67,-0.09)$~mas~yr$^{-1}$,
whereas \citet{Dia14} reported
$(\mu_\alpha\cos\delta , \mu_\delta) = (+1.06,-0.72)$~mas~yr$^{-1}$
($| \Delta \mu | = 0.74$~mas~yr$^{-1}$).

\subsubsection*{\bf Ruprecht~175:}

We see two 
peaks at $R_s \simeq 4.1$ and $R_s \simeq 6.9$~arcmin,
interestingly coinciding with the
values $R_K=4.5$ and $R_D=7.0$ (Table~\ref{tabcumulos}).
The highest is the second peak with $\eta_{max}=0.23$
which, with its associated uncertainty, yields a cluster
radius of $R_c=6.7-7.3$~arcmin.
The $119$ overdensity stars (out of $451$ stars for this
sampling) have the centroid in
$(\mu_\alpha\cos\delta , \mu_\delta) = (-1.92,-3.90)$~mas~yr$^{-1}$,
whereas \citet{Dia14}'s centroid is
$(\mu_\alpha\cos\delta , \mu_\delta) = (+2.60,-4.40)$~mas~yr$^{-1}$
($| \Delta \mu | = 0.84$~mas~yr$^{-1}$).

\subsubsection*{\bf Collinder~471:}

This is another extreme case in which we have a
very large range of $R_s$ values to be spanned
from $R_K=8.4$~arcmin to $R_D=65.0$~arcmin.
The result is also a multi-peak plot but, in this
case, values around $\sim 8.4$~arcmin are clearly
ruled out.
One of the local maxima is close to the radius
reported by D02 but, again, there is not a
clearly defined solution. For $R_s=65$~arcmin
there are $5411$ stars with $327$
in the overdensity.
The corresponding proper motion centroid is
$(\mu_\alpha\cos\delta , \mu_\delta) = (-1.57,-3.09)$~mas~yr$^{-1}$
and \citet{Dia14}'s centroid is
$(\mu_\alpha\cos\delta , \mu_\delta) = (-2.22,-3.32)$~mas~yr$^{-1}$,
($| \Delta \mu | = 0.69$~mas~yr$^{-1}$).
The shape of the proper motion distribution of
members for Collinder~471 is rather elongated
(see Fig.~\ref{figcumMPresto}). This shape
resembles the detection of substructures in
the proper motion space of the open cluster 
NGC~2548 \citep{Vic16}.
However, given that there is no a clear unique solution
and that some spurious structures might appear when 
having such a large number of field stars in the sample
($\gtrsim 5000$) the existence of such elongated
distribution is questionable.
A more detailed analysis, beyond the scope of this
study, would be necessary.

\section{Conclusions}
\label{conclusion}

In this work we have presented a method for calculating
cluster radii in a totally objective way. The MST
is used to discriminate cluster from field
in the proper motion space, and the quality of the
separation is quantified. This is done for a range
of sampling radii and the cluster radius is obtained
as the radius at which the optimal performance is
obtained. It is a different approach that does not
make use of the spatial distribution of cluster stars
(like when analysing radial density profiles).
This makes the method particularly useful for
irregular and/or poorly-populated open clusters. 

In general, the obtained cluster radius may depend
on the used astrometric catalogue, either because
the way in which the catalogue is generated can
produce some artefacts in the proper motion space,
or because of the internal precision of the data.
Here we have used UCAC4 proper motions
to determine the radii of several open clusters,
although we expect to analyse a larger sample of
cluster with precise proper motions from Gaia.
NGC~188, NGC~1647, NGC~6603 and Ruprecht~175
yielded unambiguous results. The obtained
radii for NGC~188 and NGC~1647 are
$R_c = 15.2 \pm 1.8$ and $R_c = 29.4 \pm 3.4$~arcmin,
respectively, values more or less intermediate between
the values given in D02 and K13. NGC~6603 and
Ruprecht~175 have radii of $R_c = 4.2 \pm 1.7$
and $R_c = 7.0 \pm 0.3$~arcmin, respectively,
values that are closer to D02's values than
to K13's value. Finally, both ASCC~19 and
Collinder~471 show a multi-peak behaviour
and in these cases it is not possible to be
confident about the right solution. 
It would be necessary to carry out
additional tests in oder to know
whether the multiple solutions
for these clusters are consequence
of the lack of a clear overdensity or of the
occurrence of complex patterns in their proper
motion distributions.

\section*{Acknowledgements}

We are very grateful to the anonymous referee for the
critical and constructive report that have notoriously
improved this paper.
We acknowledge financial support from Ministerio de
Econom\'{\i}a y Competitividad of Spain and FEDER funds
through grants AYA2013-40611-P and AYA2016-75931-C2-1-P.
NS has received partial financial support from Fundaci\'on
S\'eneca de la Regi\'on de Murcia (19782/PI/2014) and
Ministerio de Econom\'{\i}a y Competitividad of Spain
(FIS-2015-32456-P).
F.~L.-M. acknowledges the support by 
Funda\c{c}\~ao para a Ci\^encia e a Tecnologia (FCT)
through national funds (UID/FIS/04434/2013) and by
FEDER through COMPETE2020 (POCI-01-0145-FEDER-007672).

\bsp
\label{lastpage}
\end{document}